\documentstyle[preprint,aps,epsf]{revtex}
\preprint{\vbox{\hbox{SPhT/96-112}\hbox{hep-th/9610037}}}
\begin{document}
\title{A scenario for the $c>1$ barrier\\ in non-critical bosonic strings
}
\author{Fran\c cois David\thanks{CNRS}}
\address{Service de Physique Th\'eorique, CEA Saclay,\\
F-91191 Gif-sur-Yvette Cedex, France}
\date{Revised version - November 12, 1996}
\maketitle
\begin{abstract}

The $c\le 1$ and $c>1$ matrix models are analyzed within
large $N$ renormalization group
,
taking into account touching (or branching) interactions.
The $c<1$ modified matrix model with string exponent $\bar\gamma>0$ 
is naturally associated with an unstable fixed point, separating
the Liouville phase ($\gamma<0$) from the branched polymer phase
($\gamma=1/2$).
It is argued that at $c=1$ this multicritical fixed point and the Liouville
fixed point coalesce, and that both fixed points disappear for $c>1$.
In this picture, the critical behavior of $c>1$ matrix models
is generically that of branched polymers, but only within a scaling region
which is exponentially small when $c\to 1$.
Large crossover effects occur for $c-1$ small enough, with a $c\sim 1$
pseudo scaling which explains numerical results.

\end{abstract}
\newpage

\section{Introduction}
Matrix models have proved to be useful tools to understand several important
issues in string theory and quantum gravity
\cite{Reviews}.
It is well known that large classes of random matrix models are 
equivalent, in the so-called planar limit (number of components $N\to\infty$),
to discretized models of Euclidean 2-dimensional gravity, or equivalently
to discrete non-critical bosonic strings.
Critical points exist, which allow to take a continuum limit.
Some of these models are exactly solvable, and their continuum limit is 
2-D gravity coupled to matter with central charge $c\le 1$, or equivalently
bosonic strings in $d=c+1\le 2$ (in a linear dilaton background).
Explicit results are in agreement with other approaches, such as 
quantization in the conformal gauge (Liouville theory)
\cite{Polyakov81,David88,DisKaw89} or in the
light-cone gauge \cite{KPZ88}, or topological gravity
\cite{Witten91}.
The so-called double scaling limit 
\cite{DSL90}
allows to sum over topologies of the
2-D world sheet, and to deal with non-perturbative issues in string
theory.

One of the still poorly understood issue is the nature of the
so-called $c>1$ barrier.
No explicit solution is known for matrix models which describe
$c>1$ unitary matter coupled to 2-D gravity.
Moreover, the predictions of continuum theories for $c\le 1$ become
meaningless in the domain $c>1$. 
For instance, the KPZ formula \cite{KPZ88,David88,DisKaw89}
for the string
exponent $\gamma$ and for the scaling dimensions of matter operators
lead to unphysical complex values when $c>1$.
This $c>1$ barrier is not simply a technical problem, since it is related to
the occurrence of tachyons for bosonic strings in $d>2$.
Alternative quantizations methods for strings in $d>2$ lead only to
partial results
\cite{Klebanov}, and Liouville theory allows only to construct
consistent theories for $c=7,13,19$ \cite{Gervaisetal}, which are nevertheless 
difficult to interpret in terms of physical (non-topological) strings.

The nature of the $c>1$ phase has been investigated by numerical simulations.
The simplest case is dynamical triangulations in $d$-dimensional space
(2-D gravity coupled to $d$ massless free bosons, with $c=d>1$)
\cite{BoKaKoMi,AmDuFrOr86,DaJuKrPe87,AmDuJoTh93,AmJaTh93}.
For large values of $d$ (typically $d>10$), these studies point toward
a branched polymer phase, with a string susceptibility exponent $\gamma=1/2$
(besides ``pathological" phases with highly singular curvature defects). 
This confirms semi-rigorous 
\cite{DuFrJo84}
and more heuristic arguments 
\cite{Cates88,Seiberg90,David92}
which shows that
if $\gamma>0$ the model is ``generically" in a branched polymer phase 
with susceptibility exponent $\gamma=1/2$ .
However for smaller values of $c$ (typically $1<c<4\sim 5$) numerical
simulations point towards an intermediate phase with $0<\gamma<1/2$, without
discontinuity at $c=1$.

Another class of models consists in multiple Ising spins on dynamical
triangulations
\cite{CaKoRe92,AmDuJoTh93,BoFaHaMa94,AmTh94,KowKrz95,HaAm96,AmDuJo94}.
At the order-disorder transition point, one recovers a $c=n/2$ theory ($n$
being the number of spins).
Here also for $n>2$ but not too large $\gamma$ is found to increase smoothly
with $c$ from $0$ to $1/2$, without discontinuity at $n=2$.
For large $n$, the ordered and disordered spin phases
(with the pure gravity behavior $\gamma=-1/2$) are separated by an
intermediate branched polymer phase with $\gamma=1/2$, where the spins are already disordered
\cite{HaWh94,Harris95,HaAm96,AmDuJo94}.
The transition between the ordered  pure gravity phase and the disordered
branched polymer phase is a branching transition with $\gamma=1/3$.
A similar phenomenon occurs for $q$-states Potts models when $q$ is large
\cite{We93,AmThWe95}, and is expected for O($n$) models when $n$ is large
\cite{DuKr96} (but on flat 2-D space these models are non-critical at the
order-disorder transition for $q>4$ or $n>2$).

Analysis of $c\ge 0$ models by series expansions give qualitatively similar
results \cite{David85b,DaJuKrPe87,AmDuFr86,BreHik92}.
So if for large values of $c$ the behavior of all these models is generically
(i.e. without fine tuning) that of branched polymers, for smaller values of
$c>1$ the situation is still very unclear.
This lead several authors to conjecture that for $c>1$ but small
there exists an intermediate behavior, 
with $0<\gamma<1/2$, intermediate between the KPZ regime and the branched polymer regime \cite{KowKrz95}, while it is also argued
that very large finite size effects may occur as $c\to 1$.

Another approach is the large $N$ renormalization group (RG) introduced in
\cite{BrezinZinn92}.
The idea is that a change $N\to N+\delta N$ in the dimension of the
matrix (the string coupling constant) can be absorbed into a change in the 
couplings $g_i\to g_i+\delta g_i$ of the matrix model.
This defines a renormalization group flow, with fixed points which describe
the continuum limit.
Calculations at lowest order in perturbation theory give a picture of
the RG flow in agreement with exact results for $c<1$ models.
However the agreement is only very qualitative.
The positions of the critical points and the values for the critical exponents
are quite different from the exact results.
For some $c<1$ models one can use the equations of motion to write
new non-linear RG equations which lead to the exact critical points and
exponents
\cite{HiItNiSa93}%
, but this rely heavily on the solvability of these models, and
cannot be extended to the interesting non solvable cases.
In addition these equations are non-linear and not really in the Wilsonian
spirit.

In this article I want to show that the $c>1$ barrier can be simply understood
in this RG framework, if one takes into account the so-called touching (or
branching) interactions in the matrix models.
The scenario that I propose relies on already known results and on a few
natural assumptions.
It leads to a simple explanation of the characteristics of the $c>1$ barrier,
and of the numerical results  for $c>1$, and might be of some relevance for
higher dimensional models.

Touching (or branching) interactions generate random surfaces which touch at
isolated points,
i.e. produce a microscoping ``wormhole" connecting two world-sheets.
In matrix models they correspond to products of traces of powers of the
matrices.
For instance, the simplest matrix model with such interactions,
first introduced in \cite{Das90} ,
has for action
\begin{equation}
\label{sqrtraction}
S_N[\Phi]\ =\
 N\,{\rm tr}\left({\Phi^2\over 2}-g{\Phi^4\over 4}\right)\,-\,{x\over 2}\,
\left({\rm tr}\left({\Phi^2\over 2}\right)\right)^2
\end{equation}
where $\Phi$ is an $N\times N$ hermitean matrix.
The ${\rm Tr}(\Phi^4)$ term in the action generates the usual random
surfaces made of squares,
while the $({\rm Tr}(\Phi^2))^2$ term allows two surfaces to be glued along an
edge.
It can be solved exactly, and its phase diagram is depicted on Fig.1 .
Starting from the $x=0$, $g=1/12$ BIPZ critical point there is a critical line
which corresponds to pure 2-D gravity ($c=0$), characterized by the string
exponent $\gamma=-1/2$. 
This line ends at the end-point $x=19/241$, $g=361/5808$,
characterized by the exponent $\gamma=1/3$.
Then is becomes a branched polymer critical line,
with $\gamma=1/2$, which passes through $x=1/2$, $g=0$.
These features persist for $0<c\le 1$ matrix models
with touching interactions \cite{Kor92,AlGaBaCr93,SuTs94,GuKl94}.
When the touching coupling constant $x$  is switched on,
the 2-D gravity critical line, characterized by the exponent $\gamma\le 0$
given by the KPZ formula,
ends at a end-point, characterized by the new positive exponent
\cite{Kor92,Dur94,AmDuJo94}
\begin{equation}
\label{gammabar}
\bar\gamma\ =\ {\gamma\over \gamma-1}
\end{equation}
and becomes  for higher $x$ a branched polymer critical line, with
$\gamma=1/2$.
There is an interesting conjecture 
\cite{Klebanov94,BaDeKlSc95}
that the continuum limit of the
modified matrix model at the end point is still a Liouville theory,
but with (some) positive gravitational dressings of operators replaced by the
negative gravitational dressings. 

\section{A simple RG calculation for the $({\rm Tr}(\Phi^2))^2$ model.}
It is possible to study touching interactions with the RG approach of
\cite{BrezinZinn92}.
This is a natural idea, since $({\rm Tr})^2$ terms are naturally generated 
by the RG flow at second order in perturbation theory.
It is quite simple to apply the method of sect.~4 of \cite{BrezinZinn92}
to the model (\ref{sqrtraction}), and to obtain the 1-loop RG flow equations.

One rewrites the action (\ref{sqrtraction}) as
\begin{equation}
\label{effaction}
S_N[\Phi]\ =\
 N\,{\rm tr}\left({\Phi^2\over 2}(1-b)-g{\Phi^4\over 4}\right)\,
+\,{N^2\over 2}\,{b^2\over x}
\,+\,N^2\,\rho
\end{equation}
with $b$ an auxiliary variable, and $\rho$ an additional trivial coupling
constant, which normalizes the vacuum energy.
Starting from an $(N+1)\times (N+1)$ matrix,
one integrates over the last line $v^*$ and row $v$ of the matrix, after
rewriting the $(v^*v)^2$ term  in the action in terms of the auxiliary variable
$\sigma$ 
\begin{equation}
\label{sigmavar}
-{Ng\over 2}(v^*v)^2\ \to\ Ng\left(\sigma(v^*v)+{\sigma^2\over 2}\right)
\end{equation}
and one obtains an effective action 
$S_{\rm eff}=S_N+\Delta S$
for the remaining $N\times N$ matrix $\Phi$.
For large $N$ the variation of action is
\begin{equation}
\Delta S[\Phi]\ =\ {\rm Tr}\left({\phi^2\over 2}(1+b)-g{\Phi^4\over 4}
+{\rm Log}(1+b+g\sigma-g\Phi^2)\right)\,
+\,N\,\left( {b^2\over x}+g{\sigma^2\over 2}+2\rho\right)
\end{equation}
Assuming that $b\sim g\sim x \ll \sigma\sim 1$ we may expand the second order in
$g$, and replace $\sigma$  and $b$ by their saddle point values in
$S[\Phi]+\Delta S[\Phi]$
\begin{eqnarray}
\sigma_c\ &=&\ -(1+g+g^2)
-(x+2g-xg-g^2){1\over N}{\rm Tr}\left({\Phi^2\over2}\right)
\nonumber\\
b_c\ &=&\ -x\,{1\over N}{\rm Tr}\left({\Phi^2\over 2}\right)
\label{sigbcol}
\end{eqnarray}
We thus generate ${\rm Tr}(\Phi^2)$, ${\rm Tr}(\Phi^4)$ and $({\rm Tr}(\Phi^2))$
terms in $\Delta S$. 
Rescaling $\Phi$ in order to keep the ${\rm Tr}(\Phi^2)$ term fixed,
we find that the variation of the action amounts to a renormalization of
$g$, $x$ and $\rho$, which defines the 1-loop $\beta$-functions
%
\begin{eqnarray}
\label{betagx}
\Delta g &= -{1\over N}\beta_b(g,x)\ ;\ \beta_g =& g-6g^2-2gx\nonumber\\
\Delta x &= -{1\over N}\beta_x(g,x)\ ;\ \beta_x =& 2x-3x^2-6gx\nonumber\\
\Delta \rho &= -{1\over N}\beta_\rho(g,x,\rho)\ ;\ 
\beta_\rho =& -2\rho-{1\over 2}+{x\over 2}+{3g\over 2}+{xg\over 2}+{3g^2\over 2}
\\
\end{eqnarray}
Of course the fact that the action (\ref{sqrtraction}) remains closed under
RG transformation is true only at 1-loop.
The renormalization of $\rho$ is unimportant at that stage, and is usually
neglected.

The corresponding RG flow  in the $(x,g)$ plane is depicted on fig.~2.
The arrows goes from large $N$ to small $N$.
Besides the Gaussian attractive fixed point $(g,x)=(0,0)$, we recover the
pure gravity fixed point of \cite{BrezinZinn92} at $(g,x)=(1/6,0)$,
which has one unstable direction , as expected.
In addition, we find a purely repulsive fixed point at $(g,x)=(0,2/3)$ and
another critical fixed point at $(g,x)=(-1/6,1)$.
The interesting conclusions that we draw are:
{\it (i)} we indeed obtain a RG flow
which mixes ${\rm Tr}(\Phi^4)$ and $({\rm Tr}(\Phi^2))^2$ terms;
{\it (ii)} there are two critical points, i.e. two possible critical
behavior for the model;
and {\it (iii)} the two corresponding critical lines are separated by a
tricritical point.
However, when compared with the exact phase diagram of Fig.~1, there are
qualitative differences.
We expect that the $(0,2/3)$ fixed point corresponds to the branched polymer
critical behavior, since the $g=0$ line has an enlarged $U(N^2)$ symmetry, and
is thus stable under the RG flow.
In the first order calculation, it has two unstable directions and is thus 
multicritical, while from the exact result we find that the true
$(0,1/2)$ branched polymer critical point is critical, and should have only one
unstable direction.
One should nevertheless remember that the position of the critical points
obtained by this first order RG calculation differs from the exact positions
by $\sim 50\%$, and that the error on the critical exponents, and on the
stability of the fixed points, is very large.

\section{RG for $c<1$ models}
What lessons can be drawn from this simple calculation for $c\ne 0$ matrix
models?
We know from exact results that for $c< 1$ the phase diagram 
in the $(g,x)$ plane (where $g$ is the cosmological constant coupling and $x$
the branching coupling) is generically similar to that of Fig.~1.
If we assume that the large-$N$ RG picture stays valid, the corresponding
RG flow is schematically depicted on Fig.~3 .
There are two critical fixed points with one unstable direction,
the first one (A) corresponds to non-critical string , the second one (B)
is on the $g=0$ line and corresponds to branched polymers.
In between on the critical surface there is a multicritical point (C), with two
unstable directions, which corresponds to the modified matrix model.

The properties of the RG flow near the two fixed points A and C can be easily
deduced from the exact results.
Let us first consider the 2-D gravity fixed point A, and linearize the RG flow
around A, by mapping the couplings $(g,x,\cdots)$ into the renormalized
couplings (i.e. the scaling fields) $(t_0,t_x,\cdots)$.
Here $t_0$ is the usual cosmological constant, and $t_x$ the renormalized
branching coupling, other $t$'s correspond to possible other couplings.
In terms of the new couplings $t_i$, the singular part of the
vacuum energy $f[t]$ of the matrix model must obey the RG equation
\cite{BrezinZinn92}
\begin{equation}
\label{RGequation}
\left[N{\partial\over\partial N}+\beta_i{\partial\over\partial t_i}\right] f
\ =\ 0
\end{equation}
with the linearized $\beta$-functions
\begin{equation}
\label{betat}
\beta_i[t]\ =\ \Delta_i\,t_i+{\cal O}(t^2)
\end{equation}
which defines the scaling dimension $\Delta_i$ of the fields $t_i$.
Let us first consider the model with no branchings ($t_x=0$).
Since in the planar limit $f[t_0]$ scales like $N^2t_0^{2-\gamma}$,
using KPZ scaling \cite{KPZ88}
(we restrict ourselves to the case of unitary matter)
$\Delta_0$ is given by
\begin{equation}
\label{deltazero}
\Delta_0\ =\ {2\over \gamma-2}\ =\ -1+\sqrt{1-c\over 25-c}
\end{equation}
We can now easily deduce the dimension $\Delta_x$ of $t_x$.
Indeed, taking one derivative with respect to $t_x$ amounts to insert
one ``wormhole", i.e.
to take two planar surfaces with one puncture and to glue them at the punctures
(see Fig.~3).
This means that ${\partial\over\partial t_x}f$ scales as
${1\over N^2}\left[{\partial\over\partial t_0}f\right]^2$ and therefore that
\begin{equation}
\label{deltax}
\Delta_x\ =\ 2(\Delta_0+1)\ =\ 2\,\sqrt{1-c\over 25-c}
\end{equation}
One checks that (as long as $c<1$) $\Delta_x>0$, which means that $t_x$ is an
irrelevant coupling and that wormholes can be neglected in
the continuum limit.

Let us now consider the multicritical fixed point C, associated with the
modified matrix model.
This fixed point  has two relevant renormalized coupling, that we denote
$\bar t_0$ and $\bar t_x$.
$\bar t_0$ is the renormalized cosmological constant, already considered in
\cite{Klebanov94,BaDeKlSc95}, but we must also introduce the renormalized
branching coupling
$\bar t_x$.
In \cite{Klebanov94,BaDeKlSc95} it is shown that in the planar limit,
if $f[t_0,\cdots]$ is the
(singular part of the) vacuum energy of the ordinary matrix model in the
continuum limit, the vacuum energy $\bar f$ for the modified matrix model is
simply given by a Legendre transform
\begin{equation}
\label{barftzero}
\bar f[\bar t_0,\cdots]\ =\ {\rm extremum\ w.r.t.\ }t_0{\rm \ of\ }
\Big( f[t_0,\cdots]\,+\,t_0{\bar t_0}\Big)
\end{equation}
This general argument can easily be adapted to take into account $\bar t_x$,
and gives
\begin{equation}
\label{barftzerox}
\bar f[\bar t_0,\bar t_x,\cdots]\ =\ {\rm extremum\ w.r.t.\ }t_0{\rm \ of\ }
\Big( f[t_0,\cdots]\,+\,t_0{\bar t_0}\,+\,(t_0)^2\bar t_x\Big)
\end{equation}
(\ref{barftzero}) implies that the dimension of $\bar t_0$ is
\begin{equation}
\label{bdeltazero}
\bar\Delta_0\ =\ -2\,-\,\Delta_0\ =\ -1\,-\,\sqrt{1-c\over 25-c}
\end{equation}
(which is consistent with (\ref{gammabar})), while 
(\ref{barftzerox}) implies that the dimension of $\bar t_x$ is
\begin{equation}
\label{bdeltax}
\bar\Delta_x\ =\ -2\,-\,2\Delta_0\ =\ -2\,\sqrt{1-c\over 25-c}
\end{equation}
One checks that as long as $c<1$, $\bar\Delta_0<\bar\Delta_x<0$.
$\bar t_0$ and $\bar t_x$ are both relevant couplings, and the most relevant one
is the renormalized cosmological constant $\bar t_0$.

\section{RG for $c=1$ models}
We can now consider what happens at $c=1$. 
Then 
\begin{equation}
\label{deltacone}
\Delta_0\ =\ \bar\Delta_0\ =\ -1\qquad,\qquad\Delta_x\ =\ \bar\Delta_x\ =\ 0
\end{equation}
Assuming that there are no other fixed points in the vicinity of A and C,
the only explanation of this behavior is that the two fixed points A and C
merge at $c=1$ into a single fixed point C', with one relevant coupling $t_0$,
and one marginal coupling $t_x$.
Assuming that the branched polymer fixed point $B$ is unchanged at $c=1$,
the corresponding RG flow is depicted on Fig.~4 .
This explains in a simple way the special features of the $c=1$ theories.
Firstly, the non-analyticity of the critical exponents as $c\to 1$ is simply
reproduced by a regular variation of the RG functions as $c\to 1$.
Considering $(1-c)$ as a small parameter,
the RG functions
\begin{eqnarray}
\label{betacone}
\beta_0[t]\ &=&\ -t_0\,-\,t_0\,t_x
\nonumber\\
\beta_x[t]\ &=&\ {1-c\over 24}\,-\,t_x^2
\end{eqnarray}
simply reproduce the $c\to 1$ singularity of $\Delta_0$, $\Delta_x$,
$\bar\Delta_0$ and $\bar\Delta_x$.
Moreover this explains the logarithmic deviations to scaling
at $c=1$.
Without fine-tuning of the branching interaction, i.e. in our language for
$t_x<0$, the singular part of the free energy scales as
$f\propto t_0^2/\log(t_0)$, and at the multicritical point, i.e. for $t_x=0$,
$f\propto t_0^2\log(t_0)$
\cite{SuTs94,GuKl94}.
This is simply reproduced  with the $\beta$-functions
(\ref{betacone}), if we take into account the $\rho$ renormalization.
Indeed, the term $\beta_\rho{\partial\over\partial\rho}f=\beta_\rho$ in the RG
equation
(\ref{RGequation}) gives nothing but the additional right-hand-side $r[g]$
discussed in \cite{BrezinZinn92}\ which has therefore a standard RG
interpretation.
The correct RG equation is
\begin{equation}
\label{RGequationr}
\left[N{\partial\over\partial N}+\beta_0{\partial\over\partial t_0}
+\beta_x{\partial\over\partial t_x}
+\beta_\rho{\partial\over\partial t_\rho}
\right] f\ =\ 0
\end{equation}
and the scaling  at $c=1$ is reproduced if $\beta_0$ and $\beta_x$ are given by
(\ref{betacone}) and if we take
\begin{equation}
\label{rcone}
\beta_\rho[t]\ =\ {\rm cst}\,t_0^2\,+\,\cdots
\end{equation}

\section{RG for $c>1$ models}
This scenario for the RG flows near $c=1$ leads to a simple picture for
the $c>1$ phase.
The pair of fixed points $(A,C)$ simply disappear, and the RG flow
is schematically depicted on Fig.~5.
The only critical fixed point is then the branched polymer fixed point B,
and the whole critical line in the $(g,x)$ plane is attracted toward B.
This means that even if one starts without touching interactions ($x=0)$
they are generated by the RG and drive the model in the branched polymer phase.
This picture is in agreement with the results of the numerical simulations for
large $c$, but also explains the situation for $c$ small.
For $c-1>0$ but not too large, the pair of fixed points become a complex
conjugate pair of fixed point, not too far in the complex plane from the
real critical surface.
It is expected that there will be a strong slowing-down of the RG flow in this
region of the critical surface close to the complex fixed points (depicted
as the shaded area ${\cal A}$ on Fig.~5).
In this region, the RG flow for $c>1$ should look very much like the RG flow
close to $C'$ for $c=1$.
This implies that if one considers the matrix model without touching
interactions ($x=0$), as the cosmological constant coupling $g$ approach its
critical value $g_c$, there will be a large crossover region, up to some $g'_c$
smaller than (but close to) $g_c$, where the RG trajectory takes-off
from the critical surface in ${\cal A}$, with an effective RG eigenvalue
$\Delta_{\rm eff}\simeq -1$.
In this domain, asymptotic scaling is not reached, and one expects that
the free-energy singularity is characterized by an effective 
string susceptibility exponent
$\gamma_{\rm eff}=-2(1+1/\Delta_{\rm eff})\simeq 0$.
Only when $g'_c<g<g_c$ the true asymptotic scaling is obtained,
characterized by the branched polymer exponent $\gamma=1/2$.

If one uses the toy RG flow given by (\ref{betacone}), one can estimate the size
of the true scaling domain.
One can still simplify the RG flow equation (\ref{betacone}) into
\begin{eqnarray}
\dot t_0\ &=&\ -t_0\label{toyRG0}\\
\dot t_x\ &=&\ -\epsilon-t_x^2 %
\label{toyRGx}
\end{eqnarray}
(with $\epsilon\ =\ (c-1)/24$ and
where $\dot t$ denotes the derivative w.r.t the RG ``time", i.e. the
logarithm of the rescaling factor)
This approximation is valid for $\epsilon$ small and $t$ close to the $c=1$
fixed point $t_0=t_x=0$.
From (\ref{toyRGx}), for $\epsilon>0$ the ``time" needed to pass the
$t_x=0$ barrier (i.e. to go
from the initial condition $t_x=-\infty$ to the branched polymer fixed point at
$t_x=+\infty$) is $\simeq {\pi/\sqrt{\epsilon}}$ and from (\ref{toyRG0})
during this crossing the most relevant coupling $t_0$ inflates by a factor
$\exp(\pi/\sqrt{\epsilon})$.
Therefore if one start from $t_x\simeq-\infty$,
in order to reach
the vicinity of the BP fixed point at $t_x\simeq +\infty$ one must start
from a $t_0\sim\exp(-\pi/\sqrt{\epsilon})$.
This crude argument should nevertheless give the 
size of the true scaling domain, which is exponentially small when
$c\to 1$
\begin{equation}
\label{asympscal}
|g_c-g'_c|\ \sim\ \exp\left({-\,{{\rm constant}\over \sqrt{c-1}}}\right)
\end{equation} 
This implies that the real branched polymer scaling should be unobservable
in practice if $c-1$ is not large enough, and that one observes the
cross-over effective scaling with $\gamma\simeq 0$.
This scenario should be quite robust, and is in agreement with the numerical 
results.

\section{Application to multiple Ising spins coupled to 2D gravity}
In the above section I have considered only the cosmological constant
coupling $g$ and the branching coupling $x$, assuming that it is enough to
fine-tune $g$ (i.e to make $t_0=0$) to make the model critical.
Thus the above discussion and the phase diagram of Fig.~5 apply typically
to the non-critical string in $d>1$ space, that is to the Gaussian model
where $d$ massless free bosons are coupled to 2-D gravity.
The situation is slightly more complicated for multicritical models,
where more than one coupling have to be fine-tuned to obtain a critical theory.
A typical example is the multiple Ising spins model, which has been extensively
studied.
I shall show that the R.G. flow picture advocated in the previous section applies as well to this case, and leads to a simple and natural explanation of the numerical results.

Let me first consider the $n=1$ case.
In this discretized model, one considers a ferromagnetic Ising model
on a dynamical triangulation (in practice the spins live on the faces of the
surface).
In the model without branching the model has two couplings,
the usual string coupling $g$, and the spin temperature $T$
(the symmetry is not broken explicitly by an external magnetic field).
For fixed $T$ there is a critical coupling $g_c(T)$ where the geometry
becomes critical (infinite area).
At this $g_c(T)$, at low temperature ($0\le T <T_c$) the spins are in an
ordered phase and the fluctuations of the spins are not critical, so the
fluctuations of the geometry are that of pure gravity ($c=0$, $\gamma =-1/2$).
At high temperature ($T_c<T\le\infty$) the spins are in a disordered phase,
and the fluctuations of the metric are still that of pure gravity.
At the critical point $T_c$ the fluctuations of the spins are critical, and
it is well known that in the continuum limit one gets gravity coupled to
the $c=1/2$ minimal model (free fermions), with $\gamma=-1/3$
\cite{Kaza86}.

It is easy to add touching interactions in this model, so that one gets a model
with $g$, $T$ and the branching coupling $x$ as coupling constants.
The phase diagram for such models was already studied in
\cite{JonWhe95}.
Let me give a schematic description of the critical surface: this means that
$g$ is adjusted to its critical value $g_c(T,x)$ where the fluctuations of the
geometry are critical, and the structure of the critical surface is studied as
a function of $T$ and $x$.
For $T=0$ the spins are frozen and for $T=+\infty$ they are decoupled from the
geometry, so in both cases one recovers a phase diagram similar to that of the one matrix model as $x$ increases.
For $x$ small the system is in the pure gravity phase $\gamma=-1/2$, until
it reaches
the branching transition $\bar\gamma=1/3$, and for large $x$it is in the
branched polymer phase $\gamma=1/2$.
Now if one considers the Ising critical point ($T_c$), as $x$ is increased
it spans a critical line $T_{\rm Ising}(x)$, characterized by the $c=1/2$
gravity behavior $\gamma=-1/3$, until the branching transition
multicritical point is reached.
At this point the string exponent is $\bar\gamma_{\rm Ising}=1/4$ ,
as given by (\ref{gammabar}).
For higher $x$ the system is in the branched polymer phase, irrespective to the
value of $T$, so the pure gravity + ordered spin phase (which occurs for
small $x$ and small $T$) is separated from the branched polymer phase by
a critical line which starts from the $T=0$ branching transition point and
ends at the branching+Ising point.
Along that line the transition is just the branching transition, thus it is characterized by the branching string exponent $\bar\gamma=1/3$.
The same is true for the pure gravity + disordered spin phase
(which occurs for small $x$ and hight $T$), which is separated from the
branched polymer phase by another branching transition line, which ends also
at the branching+Ising point. 
The corresponding phase diagram (on the critical surface) is depicted on Fig.~7.
\medskip

The corresponding RG flow and fixed points on the critical surface are easy
to guess, and are depicted on Fig.~8.
{\bf A} is now the Ising fixed point, and {\bf C} the branching+Ising fixed
point.
Besides the renormalized cosmological constant $t_0$, with dimension
$\Delta_0=-6/7$, there is another
relevant coupling at {\bf A}, namely the renormalized temperature $t_2$,
which is coupled to the energy operator $\epsilon$.
Its dimension is $\Delta_2=-2/7$.
The renormalized branching coupling $t_x$ has dimension $\Delta_x=2/7$.
The unstable fixed point {\bf C} is at the end of the Ising critical line
($t_2=0$).
It has three unstable directions, corresponding to the three relevant couplings
${\bar t}_0$, ${\bar t}_2$ and ${\bar t}_x$, with respective dimensions
$\bar\Delta_0=-8/7$, $\bar\Delta_2=-2/7$, $\bar\Delta_x=-2/7$.
The $T=0$ and $T=\infty$ planes are stable under the RG flow, thus one recovers
a pure gravity f.p. {\bf A'} and a branching f.p. {\bf C'} on the zero
temperature line, and analogous f.p.'s {\bf A''} and {\bf C''} on the infinite
temperature line.
The RG flow lines going from {\bf C} to {\bf C'} and {\bf C''} correspond to
ordinary branching transitions, and separate the ordered (O) and disordered
(D) phases from the branched polymer (BP) phase.

I now consider the multiple Ising spins model, by coupling $n$ copies of the
Ising model to 2D gravity, with the same spin temperature $T$.
As long as $n<2$ one expects from the KPZ-like arguments that at the
order-disorder transition the spins are critical, and that the critical
point correspond to gravity coupled to $c=n/2$ matter.
The RG flow on the critical surface is still that of Fig.~8, with the dimensions of the couplings $t_0$, $t_x$, ${\bar t}_0$ and ${\bar t}_x$
given by (\ref{deltazero}), (\ref{deltax}), (\ref{bdeltazero}) and
(\ref{bdeltax}), and with $\Delta_2(n)={\bar\Delta}_2(n)<0$.
Assuming as above that the RG $\beta$-functions are regular as $n\to 2$, the
f.p.'s {\bf A} and {\bf C} must coalesce into a single fixed point
with two relevant directions (corresponding to $t_0$ and $t_2$), and one 
marginal direction (corresponding to $t_x$).
On the other hand as $n\to 2$ the spin fluctuations are not critical near
the zero and infinite temperature lines, so one does not expect any qualitative
change in the RG flows away from the critical $n$-Ising line.
The corresponding RG flow is schematically depicted on Fig.~9.

For $n>2$ but small the RG flow picture that one obtains is very interesting.
The pair of fixed points {\bf A} and {\bf C} disappears in the complex plane, but the RG flow along the $n=2$ critical line is almost unchanged, until one reaches the shaded region where the $n=2$ double fixed point was located.
Therefore there is now a thin ``funnel" separating the ordered (O) and
disordered (D) phases.
In this funnel one flows into the branched polymer (BP) phase.
The transition lines separating these three phases are the RG flow
trajectories going to the branching f.p.'s {\bf C'} and {\bf C''}.
This implies that the transitions are standard branching transitions with
exponent $\bar\gamma=1/3$.

The discretized models of $n$-Ising spins on dynamical triangulations have no
microscopic branching interactions, and correspond to the horizontal
$x=0$ line.
The phase diagram structure that one obtains for $n>2$ is precisely the
structure which is obtained for large $n$ \cite{HaWh94,Harris95,HaAm96}, or in the simplified model of \cite{AmDuJo94}.
The low temperature ordered phase is separated from the high temperature
disordered phase by an intermediate branched polymer phase.
The transitions are just ordinary branching transitions with $\gamma=1/3$.
Moreover at these transitions one expects that the spin fluctuations
are never critical.
The argument that I used to estimate the size of the scaling domain in
$c>1$ models can be easily adapted to estimate the width of the funnel,
that is the size $\Delta T$ (in spin temperature) of the intermediate
branched polymer phase.
One finds that it is exponentially small when $n\to 2$,
\begin{equation}
\label{deltaT}
\Delta T\ \sim\ \exp\left(-{{\rm constant}\over\sqrt{n-2}}\right)
\end{equation}
and in addition the size of the scaling domain in the most relevant
coupling $g$,
for which the branched polymer scaling and the branching transitions are
observable, still scales as (\ref{asympscal}). 
This explains why this scaling has not been observed in numerical simulation and
series analysis for moderate values of $n$, and why in practice one measures
effective scaling exponents $0<\gamma_{\rm eff}<1/2$ intermediate between the
$c=1$ scaling $\gamma=0$ and the branched polymer behavior $\gamma=1/2$.

Finally a similar picture is valid for q-states Potts models and for
the $O(n)$ interacting loops models.
For $q<4$ or $n<2$ one expects a RG flow similar to that of Fig.~8.
However at $q=4$ or $n=2$ one knows from the flat space results that the
ordinary critical fixed point will coalesce with the ordinary tricritical fixed
point, since in flat space the transition is first order for $q>4$ or $n>2$.
So for the models coupled to gravity, I expect that as $q\to 4$
or $n\to 2$ the critical f.p. {\bf A} will coalesce with an ordinary
tricritical f.p. ${\bf A}_{\rm tric}$ {\it and} with the critical branching
f.p. {\bf C}, and that they all disappear for $q>4$ or $n>2$.
The logarithmic corrections to scaling at $c=1$ will be more complicated, but
the RG flow for $q>4$ or $n>2$ will be qualitatively similar to that
of Fig.~10.
This explains the qualitative similarity between the behavior of the
multiple Ising spins models, Potts models and O($n$) models at $c>1$.
However it seems that for all these models the fluctuations of the matter
fields are never critical, and just drive some geometrical transitions of
pure gravity.

\section{Conclusions}
In conclusion, I have proposed a scenario for the $c>1$ barrier in
non-critical strings, based on the inclusion of touching interactions and
on a RG analysis.
This scenario is simple, generic, and in agreement with exact and numerical
results.
However, let me stress that it is still conjectural.
In particular, further numerical studies are needed in order to test the
scaling laws that it predicts for $c\simeq 1$.

Let me end with a few more general remarks on the role of touching interactions
in discrete gravity.
I have argued that it is necessary to include such interactions in 2-D
gravity, since they are generated by the large $N$ RG procedure
of \cite{BrezinZinn92} for matrix models.
In fact there is another equally valid reason to take them into account.
In the dynamical lattice approach of 2-D gravity, there has been several
(mostly numerical) attempts to perform real space renormalization
\cite{AmJaTh93,AmTh94,KowKrz95,JoKoKr95,BurKowKrz95,Renken94,ReCaKo95,ThCa95}.
This amounts to replace a block of neighboring cells of the random lattice by a
single, larger cell.
A problem with such procedures is that they inevitably lead to
touching points, since a small ``bottleneck" can be replaced by a single
vertex, thus generating wormhole-like configurations.
Some real-space renormalization procedures consist in removing such touching
points, when they connect a small surface (a ``baby universe")
to the parent surface \cite{JoKoKr95,BurKowKrz95}.
From the point of view that I presented here,
it is perhaps better to keep these configurations, and
to include branching interactions from the beginning in these models.

Another interesting question is to understand if touching interactions have 
a simply stringy interpretation.
For instance, I do not know if the scaling dimension $\bar\Delta_x$ of the
``wormhole operator" in the modified matrix model can be reproduced by that
of a local operator in Liouville theory.

These remarks may also apply to higher dimensional models, which have been used
to discretize 3-D and 4-D Euclidean gravity.
The numerical simulations on dynamical 3-D and 4-D triangulations show
a phase transition between a negative curvature phase and a positive curvature
phase which bears similarities with the branched polymer phase of 2-D gravity.
It is quite possible that touching interactions are needed to understand
this transition.
  
\acknowledgments
I am grateful to E. Brezin, I. Kostov and J. Zinn-Justin for useful discussions,
and to I. Kostov for his comments and his careful reading of the manuscript.
I also thank J. Ambj\o rn for his comments on the preprint, which lead to several notable improvements.

\begin{figure}
\label{fig1}
\centerline{\epsfxsize=12.truecm\epsfbox{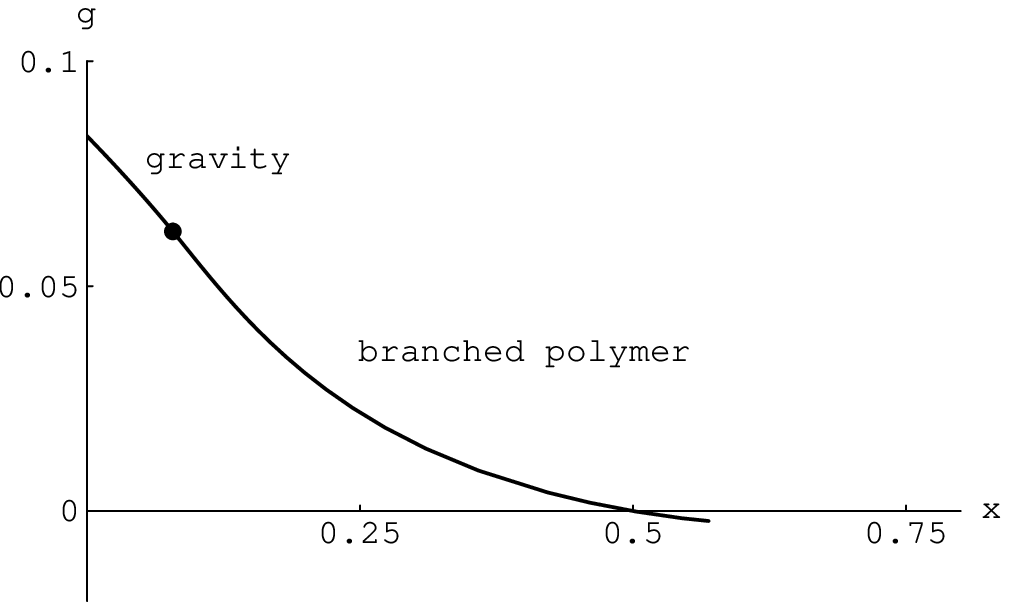}}
\caption{Exact phase diagram for the 1-matrix model with 
$({\rm Tr}(\Phi^2))^2$ interactions.}
\end{figure}

\begin{figure}
\label{fig2}
\centerline{\epsfxsize=12.truecm\epsfbox{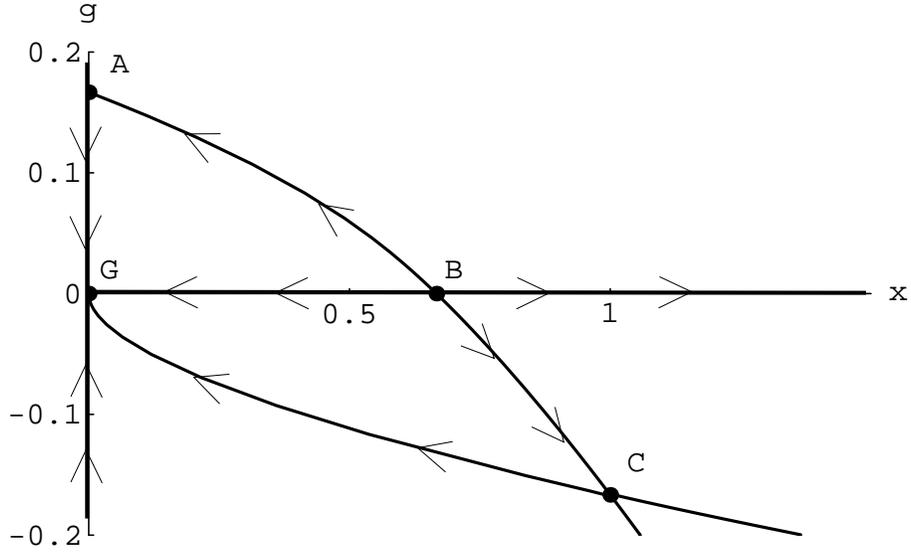}}
\caption{RG flow at first order in $x$ and $g$ for the $({\rm Tr}(\Phi^2))^2$
model.}
\end{figure}

\begin{figure}
\label{fig6}
\centerline{\epsfxsize=12.truecm\epsfbox{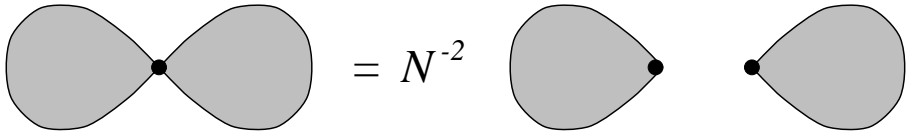}}
\caption{Inserting a wormhole}
\end{figure}

\begin{figure}
\label{fig3}
\centerline{\epsfxsize=12.truecm\epsfbox{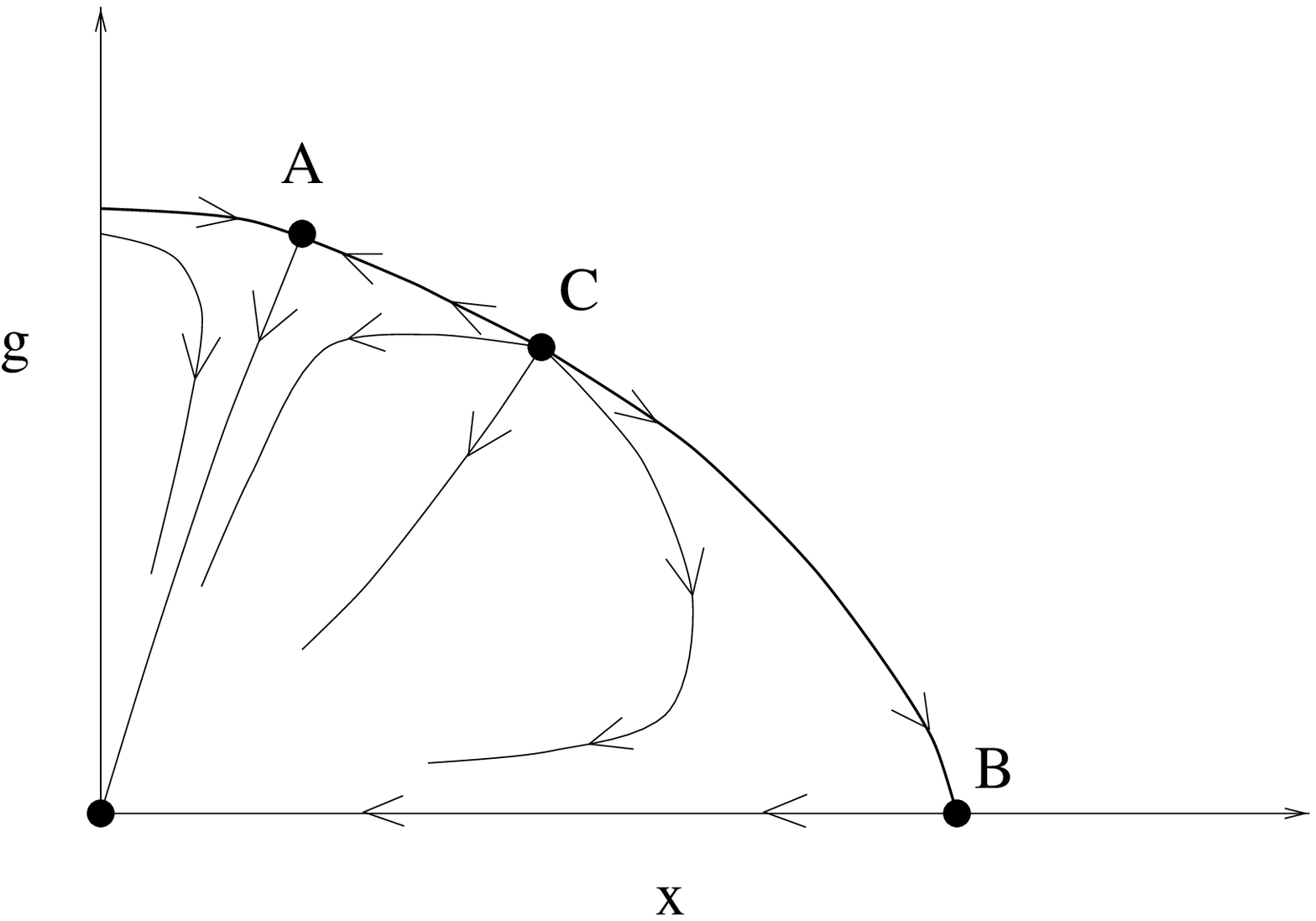}}
\caption{Schematic RG flow for $c<1$}
\end{figure}

\begin{figure}
\label{fig4}
\centerline{\epsfxsize=12.truecm\epsfbox{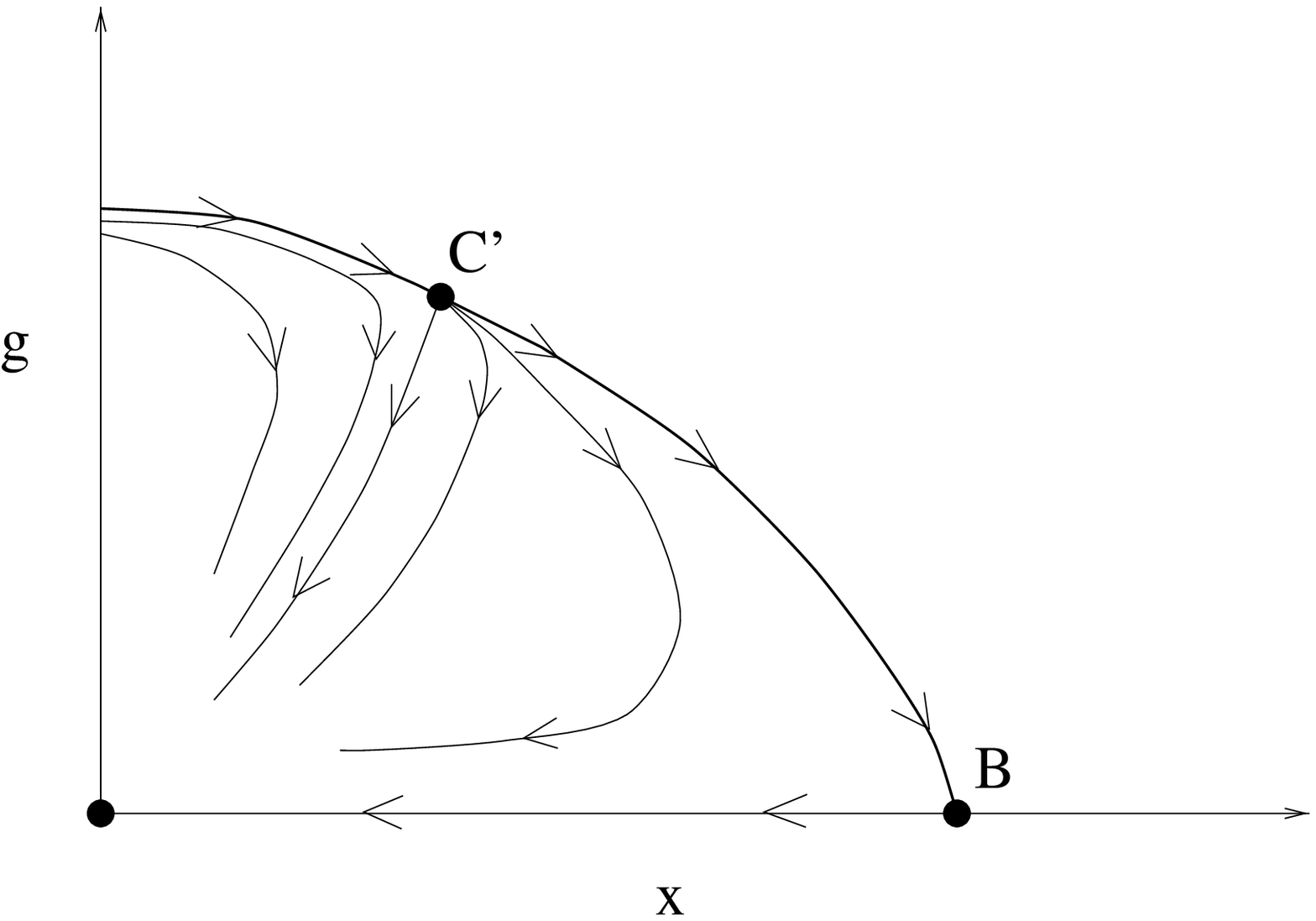}}
\caption{Schematic RG flow for $c=1$}
\end{figure}

\begin{figure}
\label{fig5}
\centerline{\epsfxsize=12.truecm\epsfbox{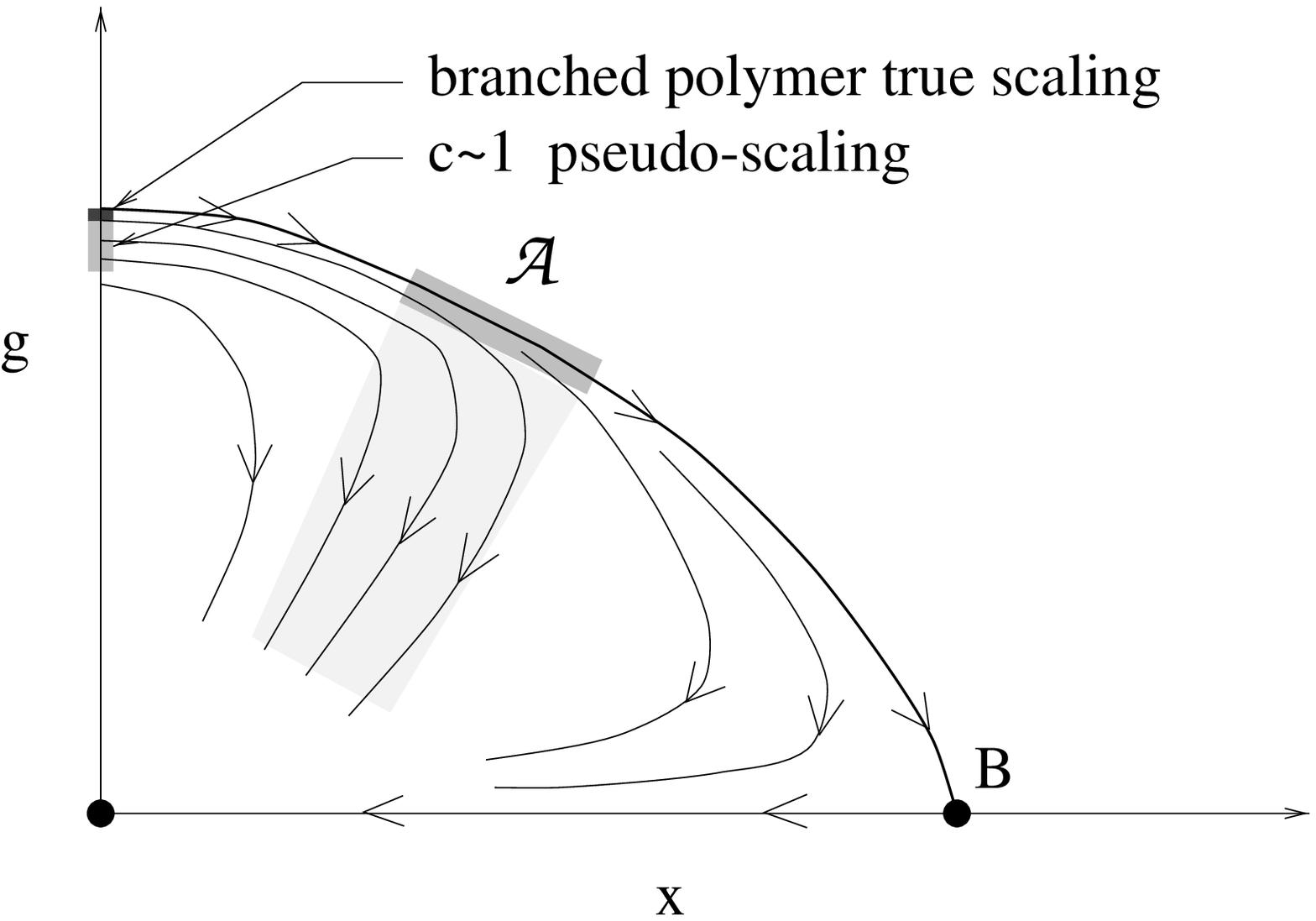}}
\caption{Schematic RG flow for $c>1$}
\end{figure}

\begin{figure}
\label{fig7}
\centerline{\epsfxsize=12.truecm\epsfbox{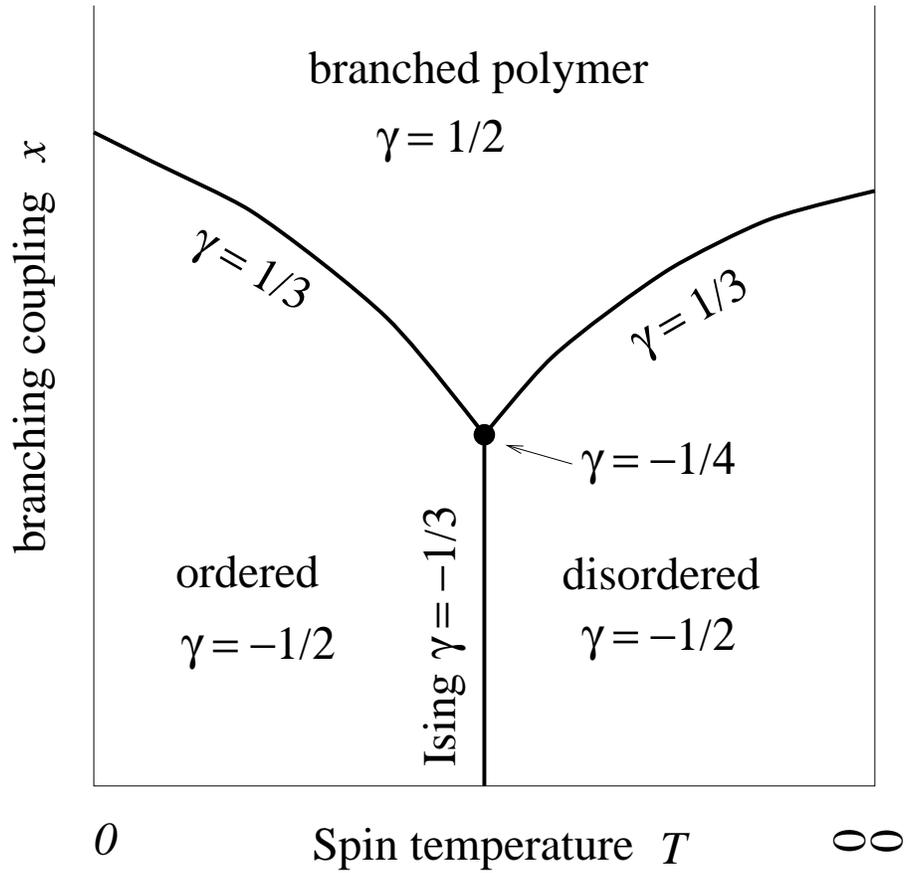}}
\caption{Schematic phase diagram for $n=1$ Ising model}
\end{figure}

\begin{figure}
\label{fig8}
\centerline{\epsfxsize=12.truecm\epsfbox{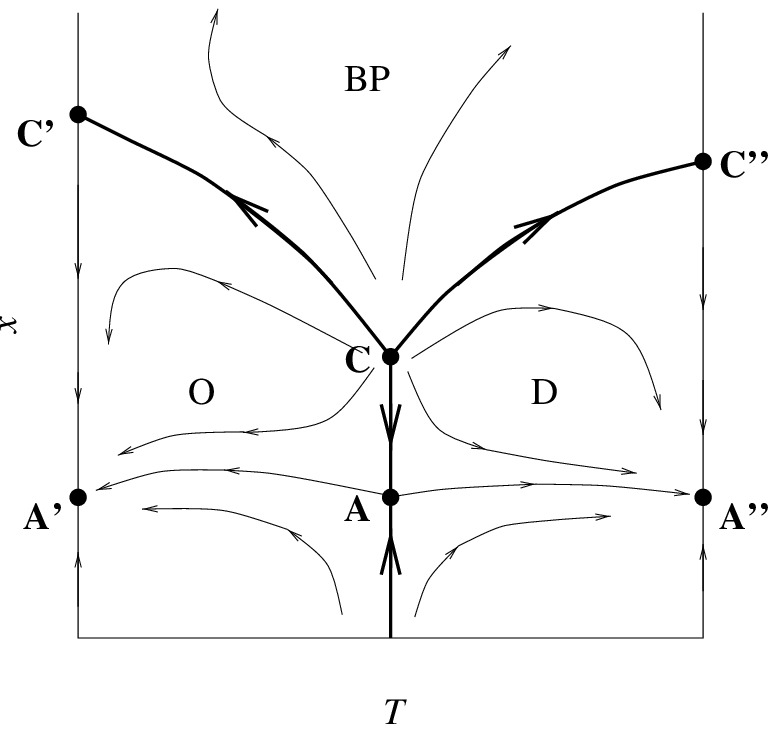}}
\caption{Schematic RG flow for $n<2$ Ising models}
\end{figure}

\begin{figure}
\label{fig9}
\centerline{\epsfxsize=12.truecm\epsfbox{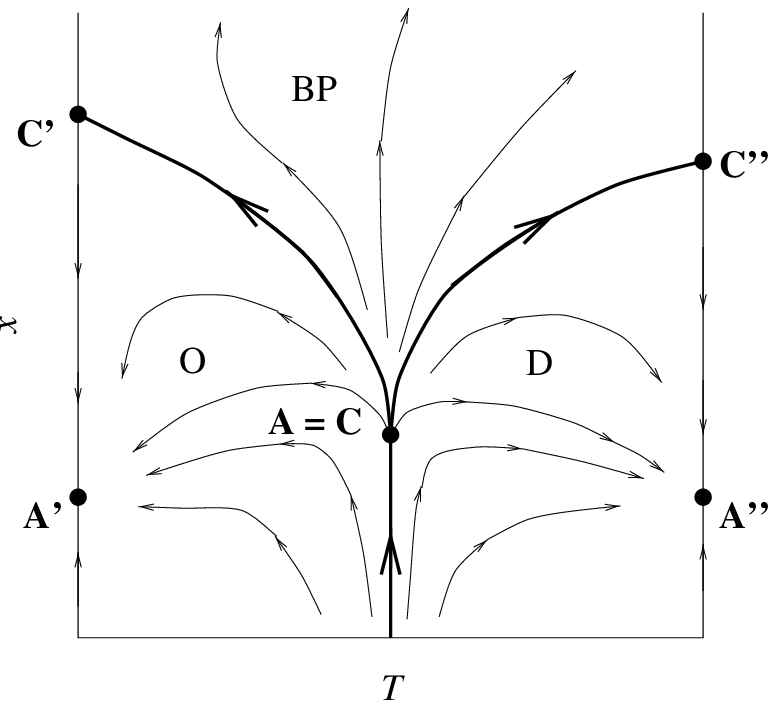}}
\caption{Schematic RG flow for $n=2$ Ising models}
\end{figure}

\begin{figure}
\label{fig10}
\centerline{\epsfxsize=12.truecm\epsfbox{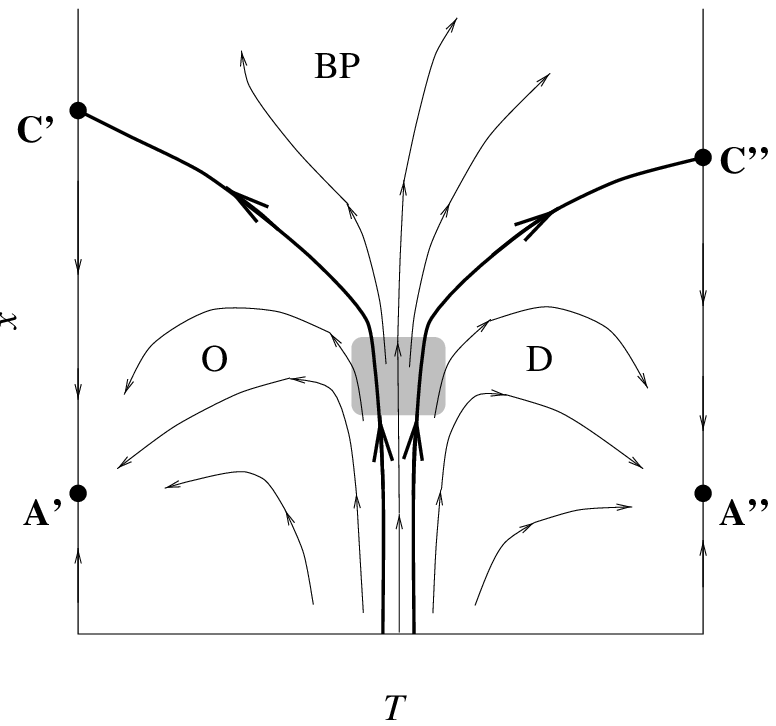}}
\caption{Schematic RG flow for $n>2$ Ising models}
\end{figure}

\end{document}